\documentclass[aps, prd, twocolumn, showpacs, nofootinbib]{revtex4}
\usepackage{amsmath}
\usepackage{graphicx}
\usepackage{dcolumn}
\usepackage{bm}
\usepackage{amssymb}
\usepackage{latexsym}
\usepackage{bm}
\usepackage{float}

\newcommand{\pa}{\partial}
\newcommand{\del}{\delta}
\newcommand{\eps}{\epsilon}

{\bf }

\newcommand{\half}{\frac{1}{2}}

\newcommand{\Del}{\Delta}

\begin{document}


\title{The fractional angular momentum realized by a cold atom}

\author{Jian Jing ${}^{a}$}
\email{jingjian@mail. buct. edu. cn}

\author{Qiu-Yue Zhang ${}^{a}$}

\author{Qing Wang ${}^{b}$}


\author{Zheng-Wen Long ${}^c$}

{}
\author{Shi-Hai Dong ${}^d$}
\email{dongsh2@yahoo. com}

\affiliation{${}^a$ Department of Physics and Electronic, School of
Science, Beijing University of Chemical Technology, Beijing 100029,
P. R. China, }

\affiliation{${}^b$ College of Physics and Technology, Xinjiang University, Urumqi 830046, P. R. China, }

\affiliation{$^{c}$ Department of Physics, GuiZhou University, GuiYang, 550025, P. R. China}

\affiliation{{ $^{d}$ Laboratorio de Informaci\'on Cu\'antica, CIDETEC, Instituto Polit\'{e}cnico Nacional, UPALM, CDMX 07700, M{e}xico. }}

\begin{abstract}

Inspired by the electromagnetic duality, we  propose an approach to realize the fractional angular momentum by using a  cold  atom which  possesses a permanent magnetic dipole moment. This atom interacts with two electric fields  and is trapped by a harmonic potential  which enable the motion of the atom to be planar and rotationally symmetric. 
We show that eigenvalues of the canonical angular momentum of the atom can take  fractional values when the atom is cooled down to its lowest kinetic energy level. The fractional part of  canonical angular momentum is dual to that of the fractional angular momenta realized by using a charged particle. Another approach of getting the fractional angular momentum is also presented. The differences between these two approaches are  investigated. 

\end{abstract}

{\pacs {03. 65. Vf, 03. 65. Pm, 03. 65. Ge}}

\maketitle


In 1984, Aharonov and Casher predicted that 
there would exist a topology phase when a neutral particle  possessing a non-vanishing magnetic dipole moment moved around a uniformly charged infinitely long filament with its direction 
paralleling to the filament \cite{AC}. It is named Aharonov-Casher (AC) effect.

In three-dimensional space, the Hamiltonian which governs the dynamics of a  neutral particle  possessing a permanent magnetic dipole moment in the background of   an electric field is given by 
\begin{equation}
H = \frac{1}{2m} ( \mathbf p - \frac{\mu}{c^2} \mathbf n \times \mathbf E)^2 + \frac{\mu \hbar}{2mc^2} \mathbf \nabla \cdot \mathbf E, \label{acha}
\end{equation}
where $m$ is the mass of the neutral particle, $\mathbf p = -i \hbar \mathbf \nabla$ is the canonical momentum,  $\mu$ is the  magnitude of the  magnetic dipole moment, $c$ is the speed of light in vacuum, $\mathbf n$ is the unit vector along the magnetic dipole moment and $\mathbf E$ is the electric field. In  AC effect setting, the electric field is produced by a uniformly charged infinitely long filament \cite{AC}. The explicit form of the electric field 
in AC effect is
\begin{equation}
\mathbf E^{AC} = \frac{\lambda}{2 \pi \eps_0 r} \mathbf e_r, \label{eac}
\end{equation}
in which $\lambda$ is  charges per unit length on the long filament, $\eps_0$ is the permittivity of vacuum, $r$ is the distance between the atom and the long filament and $\mathbf e_r$ is the unit vector along the radial direction on the plane where the atom moves.  
For  AC setting, the last term in  Hamiltonian (\ref{acha}) disappears since $\mathbf \nabla \cdot \mathbf E^{AC}=0$ for $r \neq 0$.

Hamiltonian (\ref{acha}) is the non-relativistic limit of a relativistic spin-half particle which possesses a permanent magnetic dipole moment in the background of an electromagnetic field.  
When the neutral atom moves around the uniformly charged infinitely long filament, it will receive a topology phase. The acquired topology phase is given by  
\begin{equation}
\Phi_{AC} = \frac{\mu \lambda}{\hbar c^2 \eps_0}, \label{ac}
\end{equation} 
which has been observed in the experiment \cite{EAC}.  Possible classical explanations about AC effect have been presented \cite{Boyer, APV}. Inspired by the work of Aharonov and Casher, refs. \cite{he2, whw, bakkea, bakkeb, bakkec, bakked, bakkee} studied topological phases  neutral particles would receive in various backgrounds. 

AC effect is dual to the Aharonov-Bohm (AB) effect. Over half century ago,  Aharonov and Bohm predicted that a charged particle would generate a
topology phase when it circled around a long-thin flux-carried solenoid. It is known as AB effect  \cite{AB}.  The Hamiltonian which describes a charged particle in the background of  magnetic potentials in three-dimensional space is 
\begin{equation}
H = \frac{1}{2m} (\mathbf p - q \mathbf A)^2,
\end{equation}
with  $\mathbf A$ being  magnetic potentials. Aharonov and Bohm pointed out that a  topology phase
\begin{equation}
\Phi_{AB} = \frac{q}{\hbar } \oint \mathbf A^{AB} \cdot d \mathbf l  = \frac{q \Phi}{\hbar }, \label{ab}
\end{equation}
will be generated  when the charged particle circled around the solenoid although there are no local forces exert on it. In the above expression,  $\mathbf A^{AB}$ are produced by a  long-thin flux-carried solenoid, $\Phi$ is  magnetic flux inside the solenoid and the integral is performed along a closed path which encloses the solenoid. This topology phase (\ref{ab}) has been confirmed experimentally \cite{eab, eab2}. 
The comparison 
between AB and AC effects was made in refs. \cite{Goldhaber, he}.
AB effect not only indicates that  magnetic potentials $\mathbf A$ which were introduced as auxiliaries in classical theories are observable in quantum theory, but also reveals the non-locality of  topological phases in quantum theory.

AC effect is dual to AB effect in the sense that one takes the solenoid in AB effect as a line of magnetic dipole laid end-to-end and exchanges the magnetic dipoles with the electric charge. Therefore,  for  AC effect one has a line of charges and magnetic dipoles moving around this line. The electromagnetic duality between AC and AB effects  can be understood clearly from the expressions of phases in AC and AB effects  (\ref{ac}) and (\ref{ab}), i.e.,
\begin{equation}
q \Phi \leftrightarrow \frac{\mu \lambda}{c^2 \eps_0} .   \label{dr}
\end{equation}
Actually, the electromagnetic duality has been  recognized  for quite a long time. 
The duality between AC  and AB effects can be regarded as a concrete example of electromagnetic duality. 

Another example of electromagnetic duality (\ref{dr}) was presented in \cite{es}, in which Ericsson and Sj\"{o}qvist studied a model which also describes  a neutral particle possessing a permanent magnetic dipole moment in the background of  an electric field.  Similar to the AC setting, the electric field is also along the radial direction on a plane which is perpendicular to the magnetic dipole moment. Different from AC setting, the electric field the authors applied in \cite{es}  is
\begin{equation}
\mathbf E^{ES} = \frac{\rho\, r}{2 \eps_0} \mathbf e_r, \label{ees}
\end{equation}
in which $\rho$ is the charge density.  
They found that  energy spectra of this model are analogous to the Landau levels, which are energy levels of  a planar charged particle in the background of a uniform perpendicular magnetic field. The energy  level spacing of the model  in \cite{es} is uniform, i.e.,  
\begin{equation}
\Del E = \frac{\hbar \mu \rho}{mc^2 \eps_0}=  \frac{\hbar \mu \lambda /S}{m c^2 \eps_0}, \label{gac}
\end{equation}
where $\rho =\lambda / S$ is a uniform volume charge density with the direction of $\lambda$ perpendicular to $S$.
In addition, the energy gaps of Landau levels are also uniform. They can be written as 
\begin{equation}
\Del E = \frac{\hbar q B }{m} = \frac{ \hbar q \Phi /S}{m}, \label{gll}
\end{equation}
where $B$ is the intensity of the  magnetic field,  $S$ is the area which is perpendicular to the magnetic field through which the flux $\Phi$ is measured. 
By comparing the energy gaps (\ref{gll}) for  Landau levels and Eq. (\ref{gac}) for a neutral particle which possesses a permanent magnetic dipole moment in the background of the electric field, one reproduces the duality relation (\ref{dr}). Thus, the work of \cite{es} can also be regarded as providing a theoretical approach to realize  Landau levels by a neutral particle. It may allow us to realize the quantum Hall effect by using neutral atoms and electric fields.

It is worth mentioning that the eigenvalue problem of neutral particles in various backgrounds has attracted much attention since the work of \cite{es}. In refs. \cite{bakke1, bakke1a, bakke2, bakke3, bakke4, bakke5, bakke6, bakke7}, the authors solved  energy spectra of particles  possessing non-vanishing electric or magnetic dipole moments in  the background of electromagnetic fields analytically in various configurations. 

On the other hand, with the development of  cold atomic technology, it is possible to realize some theoretical models  by cold atoms or ions. For example, Baxter \cite{baxter} proposed to realize the Chern-Simons quantum mechanics \cite{jackiw} by using a cold Rydberg atom. Ref. \cite{zhang} proposed a new approach to realize the fractional angular momenta by a cold ion. As is well-known,  eigenvalues of the angular momentum in two-dimensional  space can be fractional (values which are not quantized in the unit of $\hbar /2$) due to the Abelian nature of the rotation group \cite{wilczek, liang}. The most convenient way to realize the fractional angular momentum is to couple a charged particle with the Chern-Simons gauge field in the (2 + 1)-dimensional space-time \cite{cs1, cs2, cs3}  \footnote{This is due to the dynamical nature of the Chern-Simons gauge field and in the absence of the Maxwell term. For a review, one is recommended to \cite{forte}.}. Recently, there are renewed interests in the realization of the fractional angular momentum. In \cite{yhzhang}, the authors find that a pair of bosonic atoms immersed in a fractional quantum Hall state possesses a fractional relative angular momentum provided certain conditions are satisfied. This work was further studied in \cite{yhzhang2, lr}.  In ref. \cite{zhang}, the author considered a planar ion interacting with a uniform perpendicular magnetic field. Besides this uniform magnetic field, the ion is trapped by a harmonic potential and influenced by an Aharonov-Bohm type magnetic potential, which can be generated by a long-thin magnetic solenoid perpendicular to the plane. The dynamics of the model proposed in \cite{zhang} is described by  the Hamiltonian  (Latin indices $i, \ j$ run from 1 to 2 and the summation convention is used throughout this paper)
\begin{equation}
H= \frac{1}{2m} \Big(p_i -q (A_i + A_i ^{AB}) \Big)^2 + \half K x_i ^2 \label{hazhang}
\end{equation}
where $A_i$ are  magnetic potentials of the uniform perpendicular magnetic field and $A_i ^{AB}$ are  AB type potentials produced by the long-thin flux-carried solenoid, $\half K x_i ^2$ are the harmonic potential applied to trap the ion.

The Hamiltonian (\ref{hazhang}) can be viewed that apart from a harmonic potential, there exists a uniform magnetic field in the AB effect setting if the motion of the particle is confined on the plane perpendicular to the magnetic solenoid. Or, equivalently, besides a harmonic trapping potential, there exists an additional AB type magnetic potentials in the Landau levels setting. The eigenvalues of the canonical angular momentum of this ion are quantized, as expected. When the kinetic energy of the ion is cooled down to its lowest level, however, the author shows that the eigenvalues of the canonical angular momentum could
be fractional. The fractional part  is proportional to the magnetic flux inside the magnetic solenoid.

Both AB effect and Landau levels are related with the charged particles and magnetic potentials. Their electromagnetic dualities are all concerned with the neutral particles and electric fields. A natural question arises: can we realize the fractional angular momentum by using a neutral particle according to electromagnetic duality? In this paper, we will propose a model to realize the fractional angular momentum by using a neutral particle.

The model we considered is an atom which possesses a permanent magnetic dipole moment in the background of electric fields on a plane. The magnetic dipole moment is kept to be perpendicular to this plane. The electric fields we are applying include two parts, i.e. (\ref{eac}) and (\ref{ees}). In two dimensional space, the electric fields  (\ref{eac}) and (\ref{ees}) are 
\begin{equation}
E_i ^{AC} = \frac{\lambda x_i}{2 \pi \eps_0 r^2} \label{2ac}
\end{equation}
and
\begin{equation}
E_i ^{ES} = \frac{\rho x_i}{2 \eps_0}.  \label{2es}
\end{equation}
Besides these, the atom is trapped by a harmonic potential. 
Thus, in two-dimensional space, the Hamiltonian  takes the form
\begin{equation}
H = \frac{1}{2m}(p_i + \frac {\mu}{ c^2} \eps_{ij} E_j)^2 + \half K x_i^2 + \frac{\mu \hbar \rho }{2m c^2\eps_0 }, \label{zha3}
\end{equation}
where  $\eps_{ij}$ is the $2\times 2$ Levi-Civita matrix with  vanished diagonal elements and $\eps_{12} = - \eps_{21} =1$,  $E_i = E_i ^{AC} + E_i ^{ES}$. Here we    
have used $\pa_i E_i =  \pa_i E_i ^{ES} =\rho /\eps_0$ in the area  $(r\neq 0)$ where the atom moves. 
The Lagrangian corresponding to  Hamiltonian (\ref{zha3}) is
\begin{equation}
{ L} = \half m \dot x_i ^2 + \frac{\mu}{c^2} \eps_{ij} E_i \dot x_j - \half K x_i^2 - \frac{\mu \hbar \rho}{2m c^2\eps_0} . \label{zhla1}
\end{equation}

We pay our attention to the rotation property of the model (\ref{zhla1}). To this end, we introduce  canonical momenta with respect to variables $x_i$. They are
\begin{equation}
p_i = \frac{\pa L}{\pa \dot x_i} = m \dot x_i - \frac{\mu}{c^2} \eps_{ij} E_j. \label{zhcm1}
\end{equation}
The canonical angular momentum, by definition, is
\begin{equation}
J = \eps_{ij} x_i p_j . 
\label{amf}
\end{equation}

Applying the basic commutators
\begin{equation}
[x_i, \ x_j] = [p_i, \ p_j] =0, \quad [x_i, \ p_j] = i \hbar \del_{ij}, \label{cf1}
\end{equation}
we can verify that the canonical angular momentum (\ref{amf}) is the generator of  rotation and is conserved, i.e.,
$$[J, \ x_i] = i\eps_{ij} x_j, \ \  [J, \ p_i] = i \eps_{ij} p_j, \ [J, \ H]=0, $$
where $H$ is given in (\ref{zha3}). 

The canonical angular momentum (\ref{amf}) can also be written in the differential operator form $$J=-i\hbar \pa / \pa \varphi$$ where $\varphi$ is the  polar angle.
Obviously, eigenvalues of the canonical angular momentum (\ref{amf}) are quantized, i.e., $J_n = n \hbar, \ n=0, \pm 1, \pm 2, \cdots. $

Considering  canonical momenta (\ref{zhcm1}), we rewrite the Lagrangian (\ref{zhla1}) in the form
\begin{equation}
{ L}= \frac{ \Pi_i ^2 }{2m} + \frac {\mu}{c^2} \eps_{ij} E_i \dot x_j - \half K x_i ^2 - \frac{\mu \hbar \rho}{2m c^2\eps_0} \label{zhlar1}
\end{equation}
where $\Pi_i = m \dot x_i = p_i + \frac{\mu}{c^2} \eps_{ij} E_j$ are  kinetic momenta. Thus, the first term on the right-hand side of  above Lagrangian is  kinetic energy. 
In order to get eigenvalues of the kinetic energy algebraically, one must determine  commutators among  kinetic momenta $\Pi_i$.
After some algebraic calculations, we have
\begin{equation}
[\Pi_i, \ \Pi_j] = \frac{ i \hbar \mu \rho}{c^2 \eps_0} \eps_{ij}. \label{cmm}
\end{equation}
These commutators  remind us to introduce a pair of canonical variables $X= \sqrt{\frac{c^2 \eps_0}{\mu \rho}} \Pi_1$ and $P = \sqrt{\frac{c^2 \eps_0}{\mu \rho}} \Pi_2$ such that $[X, \ P]=1$. In terms of $X$ and $P$, we write the kinetic energy  $\frac{\Pi_i ^2}{2m}$ in Lagrangian (\ref{zhlar1}) as $\frac{\Pi_i ^2}{2m} = \frac{\mu \rho}{2 mc^2 \eps_0}(X^2 + P^2)$, which is analogous to a one-dimensional harmonic oscillator with mass $M= \frac{mc^2 \eps_0} {\mu \rho}$ and frequency
\begin{equation}
\Omega = \frac{\mu \rho}{m c^2 \eps_0}.
\end{equation}
Thus, eigenvalues of  kinetic energy can be written down directly. They are 
\begin{equation}
E_n = (n+\half) \hbar \Omega, \label{ske}
\quad
n=0, 1, 2, \cdots.
\end{equation}

Following the idea of ref. \cite{zhang}, we investigate the limit of cooling down the  kinetic energy of this atom to its lowest level. In this limit,  Lagrangian (\ref{zhlar1}) becomes 
\begin{equation}
{ L}_r= \frac{\mu}{c^2} \eps_{ij} E_i \dot x_j - \half K x_i ^2. \label{zhrl1}
\end{equation}
Obviously, in this limit, Lagrangian (\ref{zhlar1}) reduces to a form in which it does not contain  quadratic terms of velocities. Such a reduction  is similar to that applied in Chern-Simons quantum mechanics \cite{jackiw}.

Introduce  canonical momenta with respect to each $x_i$,  we get
\begin{equation}
p_i = \frac{\pa L_r}{\pa \dot x_i} = - \frac{\mu}{c^2} \eps_{ij} E_j.
\end{equation}
Since the right-hand sides do not contain velocities, they are  primary constraints in the terminology of Dirac \cite{dirac}. We label them
\begin{equation}
\phi_i ^{(0)} = p_i + \frac{\mu}{c^2} \eps_{ij} E_j \approx 0, \label{pcf}
\end{equation}
in which $'\approx'$ means  equivalent only on the constraint surface. As a result, the model (\ref{zhla1})  reduces to the one which lives in a lower dimensional phase space in this limit. 

The Hamiltonian corresponding to  Lagrangian (\ref{zhlar1}) can also be obtained by using the standard Legendre transformation. However, since the Lagrangian (\ref{zhlar1}) has been already in the first-order form, it is more convenient to read Hamiltonian from  Lagrangian (\ref{zhlar1}) directly \cite{fj}. It is 
\begin{equation}
H_r =  \half K x_i ^2.
\end{equation}
Taking into account the classical version of  basic commutators (\ref{cf1}), we obtain  Poisson brackets among  primary constraints (\ref{pcf}) as follows
\begin{equation}
\{ \phi_i ^{(0)}, \ \phi_j ^{(0)} \} = \frac{\mu \rho}{c^2 \eps_0} \eps_{ij}.  \label{pbpc}
\end{equation}
These imply that constraints (\ref{pcf}) belong to the second class and there are no secondary constraints. Because of the second class nature, the constraints (\ref{pcf}) can be used  to eliminate  redundant degrees of freedom in the reduced model (\ref{zhrl1}).

The canonical angular momentum of the model (\ref{zhrl1}) can be obtained by the definition $J = \eps_{ij} x_i p_j$. Substituting  canonical momenta into this definition, we get
\begin{equation}
J_r = \frac{\mu}{2 c^2 \eps_0} \left(\frac{\lambda}{\pi} +\rho x_i ^2\right). \label{r1am}
\end{equation}
We can also get  above angular momentum by substituting  constraints (\ref{pcf}) into $J =\eps_{ij} x_i p_j$.  Obviously, the canonical  angular momentum (\ref{r1am}) is conserved, i.e.,  $[J_r, \ H_r]=0$.

We must emphasis that  canonical angular momenta (\ref{amf}) and (\ref{r1am}) are  N\"{o}ether charges of the rotation symmetry $x_i \to x_i ^\prime = x + \delta x_i, \ \delta x_i \sim  \eps_{ij} x_j$ of the Lagrangian (\ref{zhla1}) and (\ref{zhrl1}). Therefore, the conservation of (\ref{amf}) and (\ref{r1am}) is independent of whether the parameters $\lambda$ and $\rho$ is time-dependent or not. 

In order to get  eigenvalues of the canonical angular momentum (\ref{r1am}) algebraically, let us calculate  Dirac brackets among $x_i$ and then replace them by the quantum commutators. Dirac brackets among variables $x_i$ are defined by
\begin{equation}
\{x_i, \ x_j \}_D = \{x_i, \ x_j \} - \{x_i, \ \phi_m^{(0)} \} \{\phi_m ^{(0)}, \ \phi_n ^{(0)} \}^{-1} \{\phi_n ^{(0)}, \ x_j \},
\end{equation}
where $\{ \quad, \quad \}$ is  Poisson brackets which are the classical version of  commutators (\ref{cf1}) and  $\{\phi_m ^{(0)}, \ \phi_n ^{(0)} \}^{-1}$ is the inverse matrix of (\ref{pbpc}).
A straightforward  calculation shows that
\begin{equation}
\{x_i, \ x_j \}_D = \frac{1}{i \hbar}[x_i, \ x_j]= - \frac{c^2 \eps_0}{\mu \rho} \eps_{ij}. \label{dirac}
\end{equation}
The  above commutators suggest us to introduce a pair of canonical variables $\bar X= \sqrt{\frac{\mu \rho}{c^2 \eps_0}} x_2$ and $\bar P = \sqrt{\frac{\mu \rho}{c^2 \eps_0}} x_1$ which satisfy $[\bar X, \ \bar P] = i \hbar$. Thus, the canonical angular momentum (\ref{r1am}) becomes 
\begin{equation}
J_r = \frac{\mu \lambda}{2 \pi \eps_0 c^2} + \frac{1}{2}(\bar X^2 + \bar P^2).
\end{equation}

One can easily see  that apart from a constant, the angular momentum (\ref{r1am}) is equivalent to a one-dimensional harmonic oscillator with unit  mass and frequency. The  eigenvalues of the angular momentum (\ref{r1am}) will be transparent. They are
\begin{equation}
J_n = \left(n + \half\right) \hbar + \frac{\mu\lambda}{ 2 \pi c^2 \eps_0}. \label{fam}
\end{equation}

It shows that the canonical angular momentum of the atom can  take  fractional values when the  kinetic energy of the atom is cooled down to its lowest level. Its fractional part is determined by the magnitude of  magnetic dipole moment and  parameter $\lambda$, which is  source density of the electric field $\mathbf E^{(1)}$.

It is interesting to compare our result (\ref{fam}) with that of ref.  \cite{zhang}. In ref.  \cite{zhang},  the author realized  fractional angular momenta by a cold ion and two kinds of magnetic potentials.
The eigenvalues of  the canonical angular momentum are given by \cite{zhang}
\begin{equation}
J_n = \left(n + \half\right) \hbar + \frac{q \Phi}{2 \pi}. \label{zhangr}
\end{equation}

Similar to the dualities between  AB and AC effects as well as  Landau levels 
and  eigenvalues of the model studied in ref. \cite{es},  
the fractional part of our result (\ref{fam}) is precisely dual to  (\ref{zhangr}) which is realized by using a cold ion. This means that the electromagnetic duality (\ref{dr}) holds exactly in the model  (\ref{zha3}).

In order to show that our result is reliable, we  show that the result (\ref{fam}) can also be obtained  by using  an alternative method. 

Observing the term  $\mathbf n \times \mathbf E$ in Hamiltonian (\ref{acha}) plays the same role as  magnetic vector potentials in describing a charged particle in the background of a magnetic field, we introduce the effective vector potentials and the corresponding intensity of the magnetic field. They are 
\begin{equation}
{\mathbf A}^{\rm eff} = \mathbf n \times \mathbf E 
\end{equation}
and
\begin{equation}
\mathbf B^{\rm eff} = \mathbf {\nabla \times A^{\rm eff}} .
\end{equation}
In the plane where the atom moves, the effective vector potentials  take the  form
\begin{equation}
A_i ^{\rm eff}= - \left(\frac{\lambda}{ 2 \pi \eps_0 r^2} + \frac{\rho}{2 \eps_0}\right) \eps_{ij} x_j.  \label{evmp}
\end{equation}
The corresponding  effective  magnetic field is uniform and is perpendicular to this plane 
\begin{equation}
B^{\rm eff} =  \frac{\rho}{\eps_0}.
\end{equation}
The electric field $ E_i^{AC}$ does not contribute to the effective magnetic field in the area $r \neq 0$ due to its topological nature.

Obviously, the first term on the right-hand side of (\ref{evmp}) is analogous to   magnetic potentials generated by a long-thin  solenoid and the second term is analogous to the magnetic potentials generated by a uniform perpendicular magnetic field.

Thus, in two-dimensional space the Hamiltonian (\ref{acha}) is written in the form
\begin{equation}
\tilde H = \frac{1}{2m} (p_i -\frac{\mu}{c^2} A_i ^{\rm eff})^2 + \half K x_i ^2 + \frac{\mu \hbar \rho}{2m \eps_0 c^2} \label{eha}
\end{equation}
where we  have included the harmonic scalar potential $\half K x_i ^2$. Obviously, besides a constant term, the effective model (\ref{eha}) is analogous with  the  (\ref{hazhang}). Following the same procedure we performed previously, one can get the result which is identical with (\ref{fam}).


In summary, based on the electromagnetic duality,
we provide a new approach for realizing  fractional angular momenta. Different from previous approaches which realized  fractional angular momenta by using  charged particles, we use a cold neutral atom to archive this aim. Our approach can be regarded as the electromagnetic duality of the approach proposed in \cite{zhang}. The  electromagnetic duality relation (\ref{dr}), which is found in  AB and AC effects
as well as  Landau levels and the model studied in \cite{es}, is exactly held for the results of ours and ref. \cite{zhang}.



\section*{Appendix}

In this appendix, we briefly introduce another approach to get the fractional angular momentum from the model (\ref{zha3}) in the absence of the electric field $\mathbf E^{ES}$. 

In the case of turning off the electric field $\mathbf E^{ES}$, the Lagrangian (\ref{zha3}) becomes 
\begin{equation}
\tilde L = \half m \dot x_i ^2 + \frac{\mu}{c^2} \eps_{ij} E_i ^{AC} \dot x_j - \half K x_i ^2. \label{cl}
\end{equation}
The canonical momenta are
\begin{equation}
 p_i = \frac{\pa \tilde L}{\pa \dot x_i} = m \dot x_i - \frac{\mu}{c^2} \eps_{ij} E_j ^{AC}
\end{equation}
and the  canonical angular momentum is
\begin{equation}
\tilde J = J_K +  \frac{\mu}{c^2} x_i E_i ^{AC} .\label{2am}
\end{equation}
The canonical angular momentum $\tilde J$ is conserved, i.e., $[\tilde J, \ \tilde H]=0$,  and its eigenvalues   are  quantized,  $\tilde J_n = n \hbar, \ n =0, \pm 1, \pm2, \cdots$. 

Substituting (\ref{2ac}) into the above canonical angular momentum, we get
\begin{equation}
\tilde J = J_K + \frac{\mu \lambda}{2 \pi c^2 \eps_0 }.
\end{equation}
Since the eigenvalues of the canonical angular momentum (\ref{2am}) are quantized, the eigenvalues of the kinetic angular momentum must be
\begin{equation}
{J_K}_n = n \hbar - \frac{\mu \lambda}{2 \pi c^2 \eps_0 }. \label{famk}
\end{equation}
It shows that the eigenvalues of the kinetic momentum $J_K$ are shifted by a fractional value $\frac{\mu \lambda}{2 \pi c^2 \eps_0 }$  from integers. Thus, the eigenvalues of the angular momentum $J_K$ are also fractional.

One may wonder what will happen if we cool down the kinetic energy of the atom to its lowest level in the model (\ref{cl}). A straightforward analysis shows that there will be no dynamic degrees of freedom. As a result, the fractional angular momentum will not appear.

Compared with the former approach, it seems that this approach to get the fractional angular momentum  is  more economical since there is only  electric field $\mathbf E^{AC}$ presented. However, there are some differences between them. First of all,    the quantized parts and the signs of the fractional parts of the results (\ref{fam}) and (\ref{famk}) are apparently different. Secondly, the mechanism of getting the fractional angular momenta by these two approaches are also different.  What is more,  the kinetic angular momentum  $J_K$ in (\ref{2am}) is only associated with the orbital motion. Generally, it is not conserved. To see it clearly, we notice that   $\dot {\tilde J}= \dot J_K + \frac{d}{dt} \frac{\mu \lambda}{2 \pi c^2 \eps_0}=0$. It means that $\dot J_K = - \frac{d}{dt} \frac{\mu \lambda}{2 \pi c^2 \eps_0}$.  When $\lambda$ is time-dependent, i.e., $\lambda = \lambda (t)$, the kinetic angular momentum is not conserved.  Nevertheless, as we mentioned before, the canonical  angular momentum (\ref{r1am})  is conserved,  no matter $\lambda$ is time-dependent or not. 


\section*{Acknowledgements}

The authors appreciate referees' many invaluable comments and suggestions which are great helpful to improve this manuscript.   This work is supported by NSFC with Grant No. 11465006 and partially supported by 20180677-SIP-IPN and the CONACyT under grant No. 288856-CB-2016.

\end{document}